\documentclass[prd,nofootinbib,12pt,showpacs]{revtex4}

\usepackage{amsmath,amsthm,amssymb,epsf}
\usepackage{mathbbol,bbm}
\usepackage{graphicx}
\usepackage{wasysym}
\usepackage{subfigure}
\usepackage{citesort}

\newcommand{\beqa}{\begin{eqnarray}}
\newcommand{\eeqa}{\end{eqnarray}}
\newcommand{\f}{\begin{equation}}
\newcommand{\ff}{\end{equation}}

\newcommand{\bean}{\begin{eqnarray*}}
\newcommand{\eean}{\end{eqnarray*}}
\newcommand{\ra}{\rightarrow}

\newcommand{\pa}{\partial}

\def\be{\begin{equation}} \def\ee{\end{equation}}

\begin{document}

\title{Quantum vacuum effects from boundaries of designer potentials}
\author{Tomasz Konopka}
\affiliation{ITP, Utrecht University, Utrecht 3584 CE, The Netherlands}

\preprint{ITP-UU-09/13}\preprint{SPIN-09/13}

\begin{abstract}
Vacuum energy in quantum field theory, being the sum of zero-point
energies of all field modes, is formally infinite but yet, after
regularization or renormalization, can give rise to finite
observable effects. One way of understanding how these effects
arise is to compute the vacuum energy in an idealized system such
as a large cavity divided into disjoint regions by pistons. In
this paper, this type of calculation is carried out for situations
where the potential affecting a field is not the same in all
regions of the cavity. It is shown that the observable parts of
the vacuum energy in such potentials do not fall off to zero as
the region where the potential is nontrivial becomes large. This
unusual behavior might be interesting for tests involving quantum
vacuum effects and for studies on the relation between vacuum
energy in quantum field theory and geometry.
\end{abstract}

\pacs{03.70.+k, 11.10.-z}

\maketitle

\section{Introduction}

In quantum field theory (QFT), the energy associated with the
vacuum is formally proportional to the sum of energies of all
field modes. In most situations of interest, the summation is
ultraviolet divergent. Nonetheless, because the energy spectra of
fields depend on boundary conditions, it can be argued that the
total vacuum energy does too; changing boundary conditions can
shift vacuum energy by a finite amount and produce physical and
observable effects. For example, the vacuum energy of the
electromagnetic field between two parallel conducting plates
depends on their separation and induces a macroscopic force, known
as the Casimir force, that can be measured experimentally (see
e.g. \cite{Bordag:2001qi,Milton:2004ya}). Beside the parallel
plate setup, vacuum energy has been discussed and computed in a
wide range of other situations and is known to depend subtly on
both geometry and topology
\cite{Bordag:2001qi,Milton:2004ya,Ambjorn:1981xw}.

Vacuum energy is of fundamental importance for several reasons.
Since its effects can be measured experimentally, it offers direct
verification of theoretic techniques for extracting finite
physical quantities from formally divergent expressions in QFT.
There currently seems to be an essentially sound understanding of
these issues in the laboratory context. However, since the
gravitational field in standard theory couples to the
stress-energy of matter fields and not to differences in energy,
the discrepancy between the formally divergent value of the vacuum
energy in QFT and the flatness of the observed universe is
sometimes quoted in the context of the cosmological constant
problem. Vacuum energy also appears in discussions of tests for
extra dimensions (see e.g. \cite{Milton:2004ya}).

For these and other reasons, vacuum energy has been studied in the
literature from many points of view (see e.g. the reviews
\cite{Milton:2004ya,Bordag:2001qi,Ambjorn:1981xw,Actor:1999nb} as
well as recent works involving pistons
\cite{Cavalcanti:2003tw,Hertzberg:2005pr,Edery:2006td,Lim:2008yv,Zhai:2008tf,Geyer:2008wb}).
In one approach, the summation over field mode energies is
regularized using an explicit cutoff $\Lambda$. This is a useful
approach because it reveals that the divergent contributions to
the vacuum energy are proportional to the volume and the
boundaries of the region containing the field
\cite{Cavalcanti:2003tw}. In some situations, these divergent
contributions can be removed or neutralized in a controlled
fashion \cite{Cavalcanti:2003tw}. Having parametrized them using
the scale $\Lambda$, however, tempts one to ask the question
whether their structural form, i.e. their proportionality to
volume and boundary, can be observable. Related issues have been
raised previously in discussions related to the role of boundary
conditions and materials in vacuum energy calculations
\cite{Barton:2001wd,Schaden:2006,Jaffe:2003ji}. In this paper,
such terms are shown to be observable when the field potential is
space dependent.

\medskip

The next section introduces the particular field theory studied in
this paper. It is a scalar field theory with a potential that is
quadratic in the field and that is assumed to depend on position.
The potential is used to define a cuboidal cavity to which the
field is confined, on whose boundary the field obeys Dirichlet
conditions, and within which the field has a constant mass $m$. In
this setup, the vacuum energy can be explicitly computed using the
regularization technique with cutoff $\Lambda$. In two dimensions,
the vacuum energy contains terms that are proportional to the area
and perimeter of the cavity in addition to other terms that are
either finite independently of the cutoff $\Lambda$ or vanish when
the cutoff is large. This result is then used in the context of a
potential defining two adjacent regions in a large cavity to show,
following \cite{Cavalcanti:2003tw}, that the force on a piston
separating the two regions is independent of the terms in the
vacuum energy proportional to the area and perimeter. This
calculation sets the stage for Secs. \ref{s_exotic} and
\ref{s_3d}, but also extends \cite{Cavalcanti:2003tw} by including
the field mass and other recent works on cavities with pistons
\cite{Hertzberg:2005pr,Edery:2006td,Lim:2008yv,Zhai:2008tf,Geyer:2008wb}
by discussing the effect of a soft piston on the observable force.

In Sec. \ref{s_exotic}, the potential of the scalar field is
manipulated in order to make the force on a set of pistons in a
large cavity depend on the area and perimeter of one region of the
cavity. Two distinct scenarios are described, each associated with
its own designer potential. These scenarios are highly idealized
but are nonetheless significant because they demonstrate that the
physical effects of vacuum energy do not necessarily need to
become negligible as the regions in the cavity become large. In
Sec. \ref{s_3d}, these scenarios are extended to three dimensions
and their possible observability is discussed. Section
\ref{s_conclusion} summarizes the results and discusses the
implications of the proposed scenarios on the understanding of
vacuum energy, including its role in the gravitational context.

\section{Quantum vacuum energy \label{s_pistons}}

Consider a scalar field $\phi$ in flat $d+1$ dimensional spacetime
with Lagrangian ($\hbar=c=1$) \be \label{Lgeneral} L =
\frac{1}{2}\int d^{d}x \left[(\pa_\mu \phi)(\pa^\mu \phi) -V(x)
\,\phi^2\right]. \ee The potential term is quadratic in $\phi$ and
its coefficient $V(x)$, hereafter also called the potential, is
assumed to depend on the position $x$. Field modes $\phi_{\bf{n}}$
and their energies $\epsilon_{\bf{n}}$ are found by solving the
eigenvalue equation \be \left(V(x)-\nabla^2
\right)\phi_{\bf{n}}(x) = \epsilon_{\bf{n}}^2 \phi_{\bf{n}}(x) \ee
with the appropriate boundary conditions. The potentials
considered in this paper are variations of \be \label{V0} V_0(x) =
\left\{
\begin{array}{ll} m^2 &\quad 0<x_i<a_i
\\ \infty&\quad {\mbox{elsewhere. }}
\end{array} \right. \ee This potential defines a hypercuboidal
cavity with side lengths $a_1, a_2, \ldots, a_d$ (which can be all
different). The field has mass $m$ inside the cavity; the infinite
potential outside the cavity imposes Dirichlet conditions on its
boundary and prevents the field from leaking out. For
$V(x)=V_0(x)$, field modes $\phi_{\bf{n}}(x)$ are given by
standing waves and their energies are given by \be \epsilon_{{\bf
n}}^2 =m^2+\left(\frac{\pi\, n_1}{a_1}\right)^2+\cdots
+\left(\frac{\pi \,n_d}{a_d} \right)^2, \ee with $n_i =1,
2,\ldots, \infty$. The vacuum energy, being the sum of all field
mode energies, is \be \label{E0one} E(a_1,\ldots,a_d; m) =
\frac{1}{2} \sum_{n_1=1}^\infty \cdots \sum_{n_d=1}^\infty
\epsilon_{{\bf n}}. \ee Since this expression is ultraviolet (UV)
divergent, it must be manipulated in order to extract physical
information from it.

One way to proceed is through analytic regularization. This
technique can be successfully applied in many situations and
returns a finite answer (see e.g. \cite{Ambjorn:1981xw}). However,
because the technique automatically subtracts all divergent
contributions to the vacuum energy, it also eliminates the
possibility of understanding them in detail.

A different technique that leaves one more manual control involves
introducing an explicit cutoff scale $\Lambda$ and modifying
(\ref{E0one}) into \be \label{ELambda} E_\Lambda(a_1,\ldots,a_d;
m) = \frac{1}{2} \sum_{n_1=1}^\infty \cdots \sum_{n_d=1}^\infty
\epsilon_{{\bf n}} \,D_\Lambda(\epsilon_{\bf{n}}), \ee where \be
D_\Lambda(\epsilon_{\bf{n}})=D_\Lambda\!\left(\frac{n_1}{a_d},\cdots,\frac{n_d}{a_d};m\right)
\ee is an analytic cutoff function that behaves as
$D_\Lambda(\epsilon_{\bf{n}})\sim 1$ for
$\epsilon_{\bf{n}}<\Lambda$ and $D_\Lambda(\epsilon_{\bf{n}})\sim
0$ for $\epsilon_{\bf{n}}\gg \Lambda$. This formula reduces to
(\ref{E0one}) for $\Lambda \ra \infty$ but gives a finite energy
otherwise, parametrizing the divergences in (\ref{E0one}) using
the scale $\Lambda$.

In $d=2$, when the potential $V_0(x)$ defines a rectangular cavity
with side lengths $a_1$ and $a_2$, (\ref{ELambda}) becomes \be
\label{ELambdad2} E_\Lambda(a_1,a_2; m) = \frac{1}{2}
\sum_{n_1=1}^\infty \sum_{n_2=1}^\infty
\left(m^2+\left(\frac{\pi\, n_1}{a_1}\right)^2+\left(\frac{\pi
\,n_2}{a_2} \right)^2\right)^{1/2} \!\!
D_\Lambda\!\left(\frac{n_1}{a_1},\frac{n_2}{a_2};m\right). \ee The
summations can be performed by applying the Abel-Plana formula \be
\label{AbelPlana} \sum_{n=0}^\infty F(n) = \int_0^\infty \!F(t) \,
dt \,+\, \frac{1}{2}F(0) +\, i \int_{0}^\infty \!\frac{dt}{e^{2\pi
t}-1} \left( F(it)-F(-it)\right). \ee This calculation is an
extension of the one for the massless case in
\cite{Cavalcanti:2003tw} and yields the result \be
\label{ELsummary} E_\Lambda(a_1,a_2;m) = (a_1\,a_2)\,
\mu_3(\Lambda,m)  + (a_1+a_2)\, \mu_2(\Lambda,m)  +
R_\Lambda(a_1,a_2;m). \ee In the first two terms, the functions
$\mu_3(\Lambda,m)$ and $\mu_2(\Lambda,m)$ are
\begin{align} \label{eqmu3} \mu_3(\Lambda,m) &= \frac{1}{2\pi^2} \int_0^\infty \!\!du_1\!
\int_0^\infty \!\!du_2  \left( m^2 +u_1^2+u_2^2\right)^{1/2}
D_\Lambda\left(u_1,u_2;m \right) \\ \label{eqmu2} \mu_2(\Lambda,m)
&= -\frac{1}{4\pi}\int_0^\infty \!\! du\, (m^2+u^2)^{1/2}
D_\Lambda(u,0;m). \end{align} The last term in (\ref{ELsummary}),
$R_\Lambda(a_1,a_2;m)$, is a complicated function of its
parameters
\cite{Cavalcanti:2003tw,Hertzberg:2005pr,Lim:2008yv,Edery:2006td,Zhai:2008tf}.
In most discussions of vacuum energy, it is this term that is of
most physical interest. For the present discussion, however, it is
sufficient to state that it is finite independently of the cutoff
function (its dependence on the cutoff is only $O(\Lambda^{-1})$)
and that it can be simplified to \be R_\Lambda(a_1,a_2;m) \sim w\,
m, \ee with $w$ some constant of order unity, when $\Lambda$ is
large and either $m=0$ and $a_1 \sim a_2 \gg 1$, or $m\neq 0$,
$ma_1 \gg 1$, and $ma_2 \gg 1$. That is, $R_\Lambda(a_1,a_2;m)$ is
independent of $a_1$ and $a_2$ in these regimes.

\medskip

If this setup were taken as a model for two wires of length $a_2$
separated by distance $a_1$, a physical quantity computable from
$E_\Lambda(a_1,a_2;m)$ would be the force $F$ on one of the wires
as a function of the separation distance, \be \label{Fdiverge}
\begin{split} F&= -\frac{\pa E_\Lambda(a_1,a_2;m)}{\pa a_1} \\ &=
-a_2 \, \mu_3(\Lambda,m) - \mu_2(\Lambda,m) - \frac{\pa
R_\Lambda(a_1,a_2;m)}{\pa a_1}. \end{split}\ee There are at least
two reasons why this would be a problematic result.

First, this force contains contributions that grow with the cutoff
$\Lambda$. If the cutoff were to be taken to infinity, the
resulting force would diverge and would therefore require further
manipulation. A possibility for eliminating the divergences would
be to try to subtract from (\ref{ELsummary}) the vacuum energy
associated with a region of flat space having the same shape but
without special boundary conditions imposed. This Minkowski vacuum
energy, however, would be of the form of the area term in
(\ref{ELsummary}), so the subtraction would fail to eliminate the
divergent term proportional to the perimeter; the subsequently
computed force would still contain a term that diverges for large
$\Lambda$\footnote{This kind of subtraction works well in one
dimension, but not for dimensions two or larger.}. Even if the
cutoff $\Lambda$ were assumed to be finite, perhaps related to the
Planck scale, there would still be a problem because
(\ref{Fdiverge}) would be nonzero even when $a_1\ra \infty$. Such
behavior in three dimensions would be in conflict with basic
observations.

Second, assuming that $a_1$ can be varied freely is in violations
of assumptions made in the calculations of $E_\Lambda$. More
specifically, changing $a_1$ by a finite length implies shifting
the background potential $V_0(x)$ by an infinite amount. The
problem of preserving the total background energy could be avoided
by considering joint changes in $a_1$ and $a_2$, which preserve
the area $a_1 \, a_2$. In this case, the formula (\ref{Fdiverge})
would be modified but the first issue above would remain.

\subsection{Hard Piston}

\begin{figure}
  \begin{center}
    \includegraphics[scale=1]{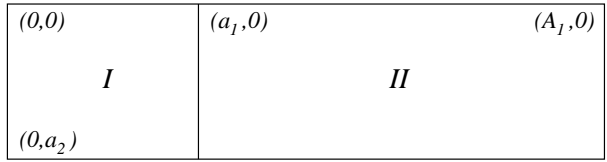}
    \caption{ A large cavity separated by a moveable piston into two
    disjoint regions. The notation (x,y) labels Cartesian coordinates of the corners of the regions. }
    \label{fig_box2}
  \end{center}
\end{figure}

An elegant resolution to these problems was proposed in
\cite{Cavalcanti:2003tw}. Instead of the small cavity with side
lengths $a_1$ and $a_2$, consider the setup shown in Fig.
\ref{fig_box2} where a large cavity with fixed side lengths $A_1$
and $a_2$ is divided by a vertical piston into two regions,
labeled $I$ and $II$, with side lengths $a_1$ by $a_2$ and
$A_1-a_1$ by $a_2$, respectively. The position of the piston is
chosen so that $A_1\gg a_1$. The potential $V(x)$ associated with
this system is \be \label{V1}
V_1(x) = \left\{ \begin{array}{ll} m^2 & \quad {\mbox{regions}}\; I, II \\
\infty&\quad {\mbox{elsewhere, }} \end{array} \right. \ee where
the piston is considered to be part of the region where the
potential is infinite. The piston is therefore ``hard'' and
enforces strict Dirichlet conditions on its surface.

The UV regulated vacuum energy $E_\Lambda^{I+II}$ for this
configuration is the combination of the vacuum energies for the
two regions, \be E_{\Lambda}^{I+II} = E_\Lambda(a_1,a_2;m) +
E_\Lambda(A_1-a_1,a_2;m).\ee Substituting the general form
(\ref{ELsummary}) for each term on the right hand side yields \be
\label{EMbox2} \begin{split} E_{\Lambda}^{I+II} &=
(A_1\, a_2) \, \mu_3(\Lambda,m) + (A_1+2a_2)\,\mu_2(\Lambda,m) \\
&\qquad + R_\Lambda(a_1,a_2;m)+R_\Lambda(A_1-a_1,a_2;m).
\end{split} \ee The result again contains terms that diverge when
$\Lambda\ra\infty$, but these terms are here independent of the
position of the piston $a_1$. The force associated with moving the
piston, $F=\pa E_\Lambda^{I+II}/\pa a_1$, thus depends only on the
regular terms $R_\Lambda$. Since these terms become independent of
$a_1$ when $A_1\gg a_1\ra\infty$, the force is then also zero
consistently with results obtained using analytical regularization
techniques \cite{Cavalcanti:2003tw}.

In this setup, moving the piston also does not change the overall
level of the background potential. This setup thus resolves both
issues discussed in association with the calculation leading to
(\ref{Fdiverge}). It does not do this by eliminating UV
divergences but rather by neutralizing them by summing
contributions to the vacuum energy from fields in two neighboring
regions. Since in laboratory situations there always exists an
outside region (region $II$) to a cavity under investigation
(region $I$), this resolution is satisfactory for all practical
purposes.

\subsection{Soft Piston}

In the above discussion, the force on a piston arises from the
terms $R_\Lambda$ in (\ref{EMbox2}), which in principle, depend on
the cutoff $\Lambda$. That these terms depend only weakly on
$\Lambda$ implies that predictions based on (\ref{EMbox2}) are
equivalent, in the practical sense, for any sufficiently large,
but not necessarily infinite, $\Lambda$. This observation suggests
that the force computable from (\ref{EMbox2}) is actually due to
low-energy effects and is independent of how very high energy
modes respond to the piston. To see this, consider again the
cavity in Fig. \ref{fig_box2} together with the potential \be
\label{V1prime}
V_{1,\,M}(x) = \left\{ \begin{array}{ll} m^2 & \quad {\mbox{regions}}\; I, \, II \\
M^2 &\quad {\mbox{piston}}
\\ \infty &\quad {\mbox{elsewhere, }} \end{array} \right. \ee
which differs from (\ref{V1}) in that it is noninfinite on the
piston - the piston is ``soft.'' It is assumed that $M \gg m$.

In this situation, the set of field modes is more complicated than
before and will not be derived in detail\footnote{The spectrum
depends, among other things, on the thickness of the piston.}.
Heuristically, however, modes with energy much smaller than $M$
should be expected to be confined by both the outer cavity walls
and the piston into the disjoint regions $I$ and $II$. In other
words, these modes obey Dirichlet conditions on the piston as well
as the cavity walls. Modes with energy much greater than $M$
should still be expected to be confined by the outer cavity walls
but should be oblivious to the presence and position of the
piston. These modes do not obey special conditions on the piston.
To a zeroth approximation, the total vacuum energy of this system
may be written as a sum of contributions from these two parts of
the spectrum, \be \label{Esoftpiston} E^{I+II}_{\Lambda,\, M} =
\sum_{\epsilon_{\bf{n}} < M} \epsilon_{\bf{n}}^{I+II} +
\sum_{\epsilon_{\bf{n}} > M} \epsilon_{\bf{n}}^{I+II}  \ee where
\begin{align} \nonumber \sum_{\epsilon_{\bf{n}} < M} \epsilon_{\bf{n}}^{I+II}
&\sim E_M(a_1,a_2;m)+E_M(A_1-a_1,a_2;m) \\ \nonumber &\sim (A_1\,
a_2) \, \mu_3(M,m) + (A_1+2a_2)\,\mu_2(M,m) \\ &\qquad +
R_M(a_1,a_2;m)+R_M(A_1-a_1,a_2;m). \label{Esoftlow} \end{align}
and
\begin{align} \nonumber
\sum_{\epsilon_{\bf{n}} > M} \epsilon_{\bf{n}}^{I+II} &\sim E_\Lambda (A_1,a_2;m) -E_M(A_1,a_2;m) \\
&\sim (A_1, a_2) \left(\mu_3(\Lambda,m) -\mu_3(M,m)\right) \nonumber \\
&\qquad +
(A_1+a_2)\left( \mu_2(\Lambda,m) -\mu_2(M,m)\right) \nonumber \\
&\qquad+ R_\Lambda(A_1,a_2;m) - R_M(A_1,a_2;m)
\end{align}
The contribution of the low energy sector is given by an
expression analogous to (\ref{EMbox2}) but with the cutoff, now
regarded as a physical one, set to $M$. The contribution of the
high energy sector is obtained by first calculating the sum of all
mode energies in the large cavity without piston (using a cutoff
$\Lambda$) and then subtracting a low energy part. The total still contains some terms that diverge in the $\Lambda\ra\infty$ limit as in the case of the hard pistons; in this sense, the soft piston does not eliminate the divergent nature of the vacuum energy.

Out of all the terms comprising (\ref{Esoftpiston}), the only ones
that depend on $a_1$ are the regular parts $R_M$ in
(\ref{Esoftlow}). When $M$ is large, these parts
depend only negligibly on this scale. Therefore, the force on the
soft piston in this cavity system is equivalent to the one
obtained in the case for the hard piston. In the present
calculation, however, the important terms are explicitly seen to
arise from low-energy modes.

The above argument is very rough. To make it more precise, one
would need to compute the energy using the exact spectrum
of the field in the cavity. In this way it should be possible to
estimate the $M$-dependent corrections to the vacuum energy and
the force on the piston. For large $M$, however, $M$-dependent corrections should be negligible and the above conclusion should be valid. In particular, there should not be important corrections to the area and perimeter terms of the vacuum energy. This is because if there were, their dependence on $a_1$ would cancel from contributions from regions $I$ and $II$. The effect of soft boundary conditions has also been discussed elsewhere in the literature, e.g. \cite{Jaffe:2003ji}.

\section{Boundary Effects \label{s_exotic}}

The calculations in the previous section reveal that divergences
in the vacuum energy are proportional to the area and perimeter of
a cavity, and that they need not be subtracted away in order to
produce reasonable results for the force on a piston. This
suggests the following question: can these nonstandard terms in
the vacuum energy ever have observable consequences?

\begin{figure}
  \begin{center}
    \includegraphics[scale=1]{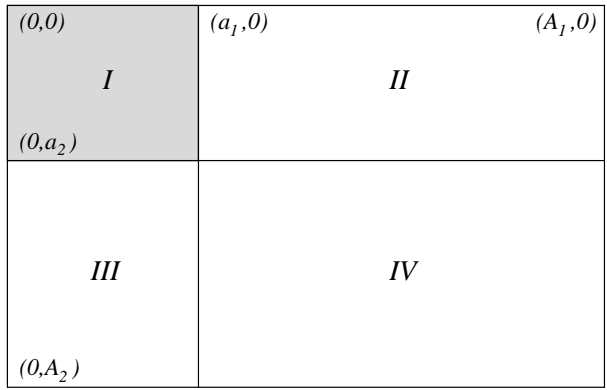}
    \caption{ A large cavity separated into four regions by two moveable pistons. \label{fig_box4}}
  \end{center}
\end{figure}

Consider the cavity configuration shown in Fig. \ref{fig_box4}.
The outer walls have side lengths $A_1$ and $A_2$. There are two
pistons, one horizontal and one vertical as shown, which divide
the cavity into four regions $I$, $II$, $III$, and $IV$. It is assumed that the two pistons can move and thereby change the size and shape of the regions. Region $I$ is shown shaded because it is assumed that the potential is different there compared to the other regions. Below are two calculations based on this cavity using  different potentials.

\subsection{Scenario 1 \label{s_exotic1}}

As a first example, consider Fig. \ref{fig_box4} together with the
potential \be V_2(x) = \left\{
\begin{array}{ll} m^2 & \quad {\mbox{region}}\; I \\ 0 &
\quad{\mbox{regions}}\; II, III, IV \\\infty&\quad
{\mbox{elsewhere. }}
\end{array} \right. \ee As before, the walls of the cavity and the
pistons are taken to belong to the region where the potential is
infinite, thus imposing Dirichlet boundary conditions there. This
is a ``designer'' potential because the field is given different
masses in different regions of space - a property not usually
considered but perhaps not completely unreasonable in the context
of theories implementing a dynamical mass-generation mechanism.

The vacuum energy for the total system is the sum of contributions
from the individual regions. Denoting the entire cavity, regions
$I+II+III+IV$ as $V$, the total vacuum energy becomes \be
\begin{split} \label{Eexotic1} E_\Lambda^{V} &= E_\Lambda(a_1,a_2;m)+
E_\Lambda(A_1-a_1,a_2;0) \\&\qquad + E_\Lambda(a_1,A_2-a_2;0)+
E_\Lambda(A_1-a_1,A_2-a_2;0)\end{split}\ee Plugging in for each of
the terms using (\ref{ELsummary}) yields \be \label{myE}
\begin{split} E_\Lambda^V &\sim (A_1\, A_2) \, \mu_3(\Lambda,0) +
(a_1 a_2) \, \left[\mu_3(\Lambda,m)-\mu_3(\Lambda,0)\right] \\
&\qquad + 2(A_1+A_2)\,\mu_2(\Lambda,0) + (a_1+a_2) \,
\left[\mu_2(\Lambda,m)-\mu_2(\Lambda,0)\right] \end{split} \ee
after ignoring all $R_\Lambda$ terms, which are negligibly small
if the lengths involved are all large. In distinction with the
calculation in the previous section, the vacuum energy now
contains terms that depend on the area and perimeter of a single
region, region $I$. The coefficients of these terms are nonzero.

Assuming \be \label{DLassumption}
D_\Lambda(u,0;m)=D_\Lambda(u,0;0),\ee the coefficient of the
perimeter term is
\begin{align} \nonumber \mu_2(\Lambda,m)-\mu_2(\Lambda,0) &=
-\frac{1}{4\pi}\int_0^\infty
du \; \left[ (m^2+u^2)^{1/2}-u\right] D_\Lambda(u,0;0) \\
\nonumber &< - \frac{1}{4\pi}\int_U^\infty du \left[
(m^2+u^2)^{1/2}-u\right] D_\Lambda(u,0;0) \\ &< - O(m^2\log
\Lambda/U) \label{sub1} \end{align} The first line is an expansion
of the $\mu_2$ functions using the definition (\ref{eqmu2}). The
integrand on this line is positive for all $u$ and thus justifies
the inequality shown next. The scale $U>0$ on the second line can
be chosen to be large so that the square root can be expanded in a
series in $m$ around $m=0$. The leading term in the integrand
becomes $m^2/2u$ and thus the integral depends logarithmically on
the cutoff. A similar analysis for the other coefficient gives \be
\label{sub2} \mu_3(\Lambda,m)-\mu_3(\Lambda,0) > O(m^2 \Lambda).
\ee

As long as condition (\ref{DLassumption}) holds, the scalings and
signs in (\ref{sub1}) and (\ref{sub2}) are general and cannot be
removed by choosing a special form for the cutoff function. If
(\ref{DLassumption}) does not hold, i.e. if the cutoff function is
allowed to depend on $m$, then the mass $m$ can play a subtle role
in the integrals and the dependence of the coefficients on the
cutoff may in some cases be removed.\footnote{I would like to
thank Jan Ambjorn for pointing this out.} However, even by
tweaking the cutoff function in this manner, neither (\ref{sub1})
or (\ref{sub2}) can be made to vanish completely.

While both the area and perimeter terms are in principle
observable, suppose that a constraint is imposed keeping the area
$a_1 a_2$ of region $I$ fixed. In this case the term proportional
to the area would be unobservable. However, since the potential is
fluid (by assumption), moving the pistons would change the vacuum
energy by the term proportional to the perimeter, and this would
produce a nonzero force. The sign of the perimeter term implies
that the vacuum energy, given a fixed area $a_1 a_2$, decreases as
$a_1$ and $a_2$ become more unequal.

Suppose that the hard pistons were replaced by soft ones so that
modes with energy larger than $M$ would need to satisfy Dirichlet
conditions on the outer cavity walls but not on the piston. One
aspect of this modification would be that the form
(\ref{Eexotic1}) would be an accurate description of the
contribution of only the low-energy modes. The coefficients of the
area and perimeter terms would thus scale with $M$ rather than
$\Lambda$. High energy modes should be relatively unaffected by
the positions of the pistons and so might not produce important
$\Lambda$-dependent terms either. The magnitude of the boundary
contribution would thus not depend on the cutoff and the soft
piston could be said to regularize the observable terms in the
vacuum energy. It is important to note, however, that the source
of the observable terms would not lie in the precise nature of the
soft pistons but rather on the different masses of the scalar
field in the various regions of the cavity. In any case, since
soft boundary conditions can sometimes lead to subtle effects
\cite{Jaffe:2003ji,Schaden:2006}, this issue should be
investigated in more detail. It is not unreasonable to suggest
that a more realistic setup than the one described in this section
might lead to vacuum energy terms that are independent of the
cutoff $\Lambda$.

\subsection{Scenario 2 \label{s_exotic2}}

As a second example, consider Fig. \ref{fig_box4} and the
potential
\be V_3(x) = \left\{ \begin{array}{ll} m^2(\epsilon_{\bf{n}}) & \quad {\mbox{region}}\; I \\
0 & \quad{\mbox{regions}}\; II, III, IV
\\\infty&\quad {\mbox{elsewhere. }} \end{array} \right. \ee Here,
the mass of the field is not only different in the various regions
of the cavity, but it is also dependent on the mode energy. The
function $m(\epsilon_{\bf{n}})$ is assumed to be behave as
$m(\epsilon_{\bf{n}})\sim 0$ for $\epsilon_{\bf{n}} \gg M$ and
$m(\epsilon_{\bf{n}}) \gg M$ for $\epsilon_{\bf{n}}<M$.
Effectively, this makes region $I$ transparent to high energy
modes but not to low energy ones.

The vacuum energy for this cavity, using again the notation
$I+II+III+IV=V$, is \be \begin{split} E_\Lambda^V &=
\left[E_\Lambda(a_1,a_2;0) -G_M(a_1,a_2;0)\right]+
E_\Lambda(A_1-a_1,a_2;0)
\\&\qquad + E_\Lambda(a_1,A_2-a_2;0)+
E_\Lambda(A_1-a_1,A_2-a_2;0),
\end{split} \ee where $G_M$ describes the effect of the
potential on low-energy field modes in region $I$. Without this
term, the sum of the remaining pieces produces a quantity in which
all area and perimeter terms are independent of the piston
positions $a_1$ and $a_2$. Any dependence of the vacuum energy on
the piston positions is therefore encoded in the term $G_M$. If
the mass function $m(\epsilon_{\bf{n}})$ is sufficiently sharp to
eliminate all modes with energies up to the scale $M$, \be
m(\epsilon_{\bf{n}}) = \left\{ \begin{array}{ll} 0 & \quad
{\mbox{if }}\; \epsilon_{\bf{n}} > M
\\\infty&\quad {\mbox{if }}\, \epsilon_{\bf{n}} \le M, \end{array} \right.
\ee then $G_M$ will have the same form as (\ref{ELsummary}), \be
G_M(a_1,a_2;0) \sim (a_1 \, a_2)\, \mu_3(M,0) + (a_1+a_2)\,
\mu_2(M,0) + R_M(a_1,a_2;0). \ee If the area of region $I$ is
constrained to be fixed, the first term is unobservable. The last
term can be made negligibly small. That leaves only the boundary
term. Interestingly, this term is now not divergent but is
dependent on the scale $M$ associated with the nontrivial
potential in region $I$. The result is a quantum vacuum effect
that depends on the boundary of a region and a noninfinite energy
scale. Since the sign of the boundary term is here opposite to
that in the previous scenario, the vacuum energy is here lowest
when $a_1 \sim a_2$.

In this setup, since the observable terms in the vacuum energy are
independent of $\Lambda$, replacing the hard pistons by soft ones
would not change the core argument and result.

\section{Three Dimensions \label{s_3d}}

All the above arguments can be extended to three dimensions. The
vacuum energy in a cuboidal cavity with side lengths $a_1,$ $a_2$,
and $a_3$ and potential $V_0$ in (\ref{V0}) is \be \label{EL3d}
\begin{split} E_\Lambda(a_1,a_2,a_3;m) &= (a_1\, a_2\, a_3) \,
\mu_4(\Lambda,m) -\frac{1}{2}\, (a_1 \,a_2 + a_1\,a_3+a_2\,a_3) \, \mu_3(\Lambda,m) \\
&\qquad -\frac{1}{2}\,(a_1+a_2+a_3) \,\mu_2(\Lambda,m) + \!
R_\Lambda(a_1,a_2,a_3;m). \end{split} \ee Here
$R_\Lambda(a_1,a_2,a_3;m)$ is a regular function that, like
$R_\Lambda(a_1,a_2;m)$ in two dimensions, becomes a constant in
the limits $\Lambda\ra\infty$ and $a_1,a_2,a_3 \ra \infty$ (see
e.g. \cite{Ambjorn:1981xw,Lim:2008yv,Edery:2006td}). The functions
$\mu_3(\Lambda,m)$ and $\mu_2(\Lambda,m)$ are the same as in
(\ref{eqmu3}) and (\ref{eqmu2}), respectively, and
$\mu_4(\Lambda,m)$ is \be \mu_4(\Lambda,m) = \frac{1}{2\pi^3}
\int_0^\infty \!\!du_1 \int_0^\infty \!\!du_2 \int_0^\infty
\!\!du_3 \left( m^2+u_1^2+u_2^2+u_3^2\right)^{1/2}
D_\Lambda(u_1,u_2,u_3). \ee Thus, the divergent terms are here
proportional to the volume $V=a_1\,a_2\,a_3$, surface area
$S=a_1\, a_2 + a_1\,a_3+a_2\, a_3 $, and total edge length
$L=a_1+a_2+a_3$ of the cavity.

The three-dimensional analog of Fig. \ref{fig_box4} is a cavity
setup with one special region and $2^3-1=7$ trivial regions,
separated by three moveable pistons. Assuming again that the
volume of the nontrivial region is constant, the volume
contribution in total vacuum energy cannot change when the pistons
move. But shifts in the vacuum energy can be achieved through
changes of the surface area $S$ and total edge length $L$. In a
system analogous to the one in Sec. \ref{s_exotic1}, changes in
vacuum energy would take the form \be \label{DeltaE1}\Delta
E_\Lambda \sim -(\Delta S) \, O(\Lambda m^2) + (\Delta L) \, O(m^2
\log \Lambda/m). \ee A system like the one in Sec. \ref{s_exotic2}
would produce vacuum energy changes on the order of \be
\label{DeltaE2} \Delta E_\Lambda \sim (\Delta S) \, O(M^3) -
(\Delta L)\, O(M^2). \ee Contributions from the regular parts
$R_\Lambda$ are omitted in both formulas.

Consider now the following estimates for the magnitudes of these
changes in vacuum energy. Consider first (\ref{DeltaE2}) and
assume that $M$ is associated with the atomic scale, $M\sim 10^3
eV$. Changing the perimeter of the filled region by one square
meter and the edge lengths by one meter would give $\Delta E \sim
10^{23}eV$ and $\Delta E \sim 10^{13}eV$, respectively. Consider next (\ref{DeltaE1}) with $\Lambda\sim 10^{28} eV$ and
$m\sim 10^{6}eV$, so that the cutoff is identified with the Planck
scale and $m$ with the masses of elementary particles. Changing
the perimeter of the filled region by a square meter and the edge
lengths by one meter would now give $\Delta E \sim 10^{54} eV$ and
$\Delta E \sim 10^{20} eV$, respectively.

In both cases, the described changes in vacuum energy are roughly
comparable to thermal effects (given by $kT$ times a suitable
macroscopic number of degrees of freedom). They are therefore not
automatically ruled out by observations and may in fact play
interesting roles in the physics of the described situations. The
difficulties for observing these effects directly, however, come
in producing the required potentials and in adjusting the shapes
of the nontrivial regions in a controlled fashion.

\section{Conclusion \label{s_conclusion}}

In summary, this paper considers vacuum energy associated with a
quantum scalar field confined to various cuboidal cavities. These
simple geometries allow one to compute the vacuum energies
explicitly and regularize their divergences using a cutoff
$\Lambda$. In general, these divergences are proportional to the
size (volume) and shape (boundary) of the cavities. In
calculations extending \cite{Cavalcanti:2003tw}, it is shown that
if a cuboidal cavity is divided into distinct regions by pistons,
the forces on the pistons are independent of the volume and
boundary terms if the mass of the field is the same in all
regions. The forces are also argued to be independent of whether
the pistons are infinitely hard or soft. In Secs. \ref{s_exotic}
and Sec. \ref{s_3d}, however, it is shown that boundary terms in
the vacuum energy can lead to observable effects under certain
circumstances. Two distinct scenarios are proposed.

In the first scenario, introduced in Sec. \ref{s_exotic1}, the
field is assumed to be massive in one region of a cavity, and
massless in other regions. The mass discrepancy leaves an
observable term in the vacuum energy that is proportional to the
boundary of the region where the field is massive. This term also
depends non-negligibly on the cutoff $\Lambda$ - it diverges if
the cutoff is taken to infinity. This could be seen as a pathology
or an opportunity, depending on the point of view. In any case, it
is unclear whether the cutoff dependence is an artifact of using a
particular form of the cutoff function or assuming,
unrealistically, that the pistons separating the two regions of
space are infinitely hard. It is possible that using soft pistons
in that calculation may yield an effect that depends on the mass
$m$ of the field or on another finite, intermediate scale that
reflects the hardness or softness of the pistons. This issue is
related to discussions of boundary conditions in other vacuum
energy systems \cite{Jaffe:2003ji,Schaden:2006}.

In the second scenario, described in Sec. \ref{s_exotic2}, one
region of a cavity has a negligible potential for field modes with
energy above a threshold $M$ and an effectively infinite potential
for field modes with energy below $M$. The observable part of the
vacuum energy in this case again contains a term proportional to
the volume and boundary of the special region. The mass scale
associated with the effect in this case is $M$. Curiously, the
sign of the boundary effect is opposite to that in the first
scenario.

These effects are interesting for both theoretical and
observational reasons. On the theoretical side, the scenarios
described test understanding of regularization and renormalization
methods in QFT. In particular, a naive computation of vacuum
energy in the first scenario with hard pistons, Sec.
\ref{s_exotic1}, using analytic regularization techniques would
not predict a boundary effect. The described effect thus
differentiates the cutoff and analytic regularization approaches
and offers a way to determine which is the more correct way of
understanding vacuum energy in QFT. (The boundary effect in the
second scenario, Sec. \ref{s_exotic2}, could be argued to arise
also within the analytic regularization scheme.)

It is also interesting to compare the arguments and results of the
two scenarios with other work which has shown that binding
energies of bodies composed of a granular material are
proportional to their volumes and boundaries \cite{Barton:2001wd}.
While the scaling of those effects is qualitatively similar to
those in the two scenarios in Sec. \ref{s_exotic}, there are
important differences between the two sets of calculations. One
important difference is in the style of calculation. The
calculations (with hard pistons) in Sec. \ref{s_exotic} are exact
and do not depend on any subtraction prescription. They therefore
emphasize the source of the observable volume and boundary terms
as due to the nontrivial potential. Another important difference
between the two approaches is that in the first scenario in Sec.
\ref{s_exotic1}, the observable effects arise due to mass
differences in distinct regions of space which can arise due to
dynamics of a single field and can therefore arise in vacuum
without interactions with any granular materials.

These effects might also have interesting implications for issues
in quantum gravity. The idea that the Planck scale might provide a
real UV cutoff for quantum field theory appears in many forms in
the literature (see e.g. \cite{Garay} for general arguments and
e.g. \cite{Physicalcutoff1,Physicalcutoff2,Physicalcutoff3,Physicalcutoff4}
for a few recent approaches). The present discussion suggests that
quantum gravity models should take into account
$\Lambda$-dependent contributions to the vacuum energy
proportional to the boundary as well as the volume of the
observable universe. That is, they should not only account for why
the volume contribution to the vacuum energy does not generate a
large cosmological constant, but should also explain the role of
the boundary terms (for possible consequences of the boundary
terms on geometry, see \cite{KonopkaDICE}.)

On the observational side, the boundary terms are interesting
because their magnitudes increase with the size of a region. This
is an interesting behavior for vacuum energy whose effects, apart in the case of materials \cite{Barton:2001wd}, are
usually inversely proportional to the separation between
boundaries and therefore vanish for large systems. Also, in
contrast with other works where boundaries have been found to play
a role (e.g. \cite{Wagner:2008qq}), the effects described in this
paper are dominant and not corrections. Simple estimates of the
magnitudes of the boundary energies in three dimensions in Sec.
\ref{s_3d} show that they can be on the order of thermal energies
and hence that they are not immediately ruled out by existing
observations. Testing for the boundary energies should therefore
be interesting as a matter of principle. In practice, however,
observation might be difficult due to the peculiar potentials
required and due to the necessity of changing the shape and volume of regions of space in a precise and controlled fashion.

\begin{acknowledgments}
I would like to thank Jan Ambjorn and Renate Loll for useful
discussions.
\end{acknowledgments}

\end{document}